\documentclass[aps,prb,twocolumn]{revtex4}
\usepackage{epsfig}

\begin{document}

\title{Hot water can freeze faster than cold?!?}

\author{Monwhea Jeng}

\email{mjeng@siue.edu}

\affiliation{
Physics Department, Box 1654,
Southern Illinois University Edwardsville,
Edwardsville, IL, 62025}

\begin{abstract}
We review the Mpemba effect,
where intially hot water
freezes faster than initially cold water. While the effect
appears impossible at first sight, it has been seen in numerous
experiments, was reported on by Aristotle, Francis Bacon, and
Descartes, and has been well-known as
folklore around the world.
It has a rich and fascinating history, which
culminates in the dramatic story of the
secondary school student, Erasto Mpemba, who
reintroduced the effect
to the twentieth century scientific community.
The phenomenon, while simple to describe, is
deceptively complex, and illustrates
numerous important
issues about the scientific method:
the role of skepticism in scientific inquiry, the
influence of theory on experiment and observation,
the need for precision in the statement of a scientific
hypothesis, and the nature of falsifiability.
We survey proposed theoretical mechanisms for the 
Mpemba effect,
and the results of modern experiments on 
the phenomenon.
Studies of the observation that hot water
pipes are more likely to burst than cold water pipes
are also described.

\end{abstract}

\maketitle


\section{Introduction}
\label{sec:introduction}

The Mpemba effect occurs when two
bodies of water, identical in every way, except that
one is at a higher temperature than the other, are
exposed to identical subzero surroundings, and the 
initially hotter water freezes first.
The effect appears theoretically impossible at first
sight.
Nevertheless, it has been observed
in numerous experiments, and we will see that it
is in fact possible.

Readers who are quite certain
that the effect  is forbidden by the laws of
thermodynamics should pause for a
moment to do two things. First,
to try to explain, as precisely as possible, 
why it is impossible. And second, to 
decide how to respond to a non-scientist who
insists that they have
observed the Mpemba effect
in a home experiment.
Whether or not the effect is real, careful consideration
of these tasks will bring up a wealth of important issues
about the scientific method, that can be understood and
discussed by students without any knowledge of advanced
physics.

In section~\ref{sec:prehistory}, we describe early
observations and experiments on this phenomenon.
The effect was discussed by, among others, 
Aristotle, Roger Bacon, Francis Bacon, and Descartes.
The effect was repeatedly discussed in 
support of
an {\it incorrect} theory of heat, 
and appears to have been forgotten once more
modern conceptions of heat, which it {\it appears}
to contradict, were developed.
In fact, we will see that Kuhn
incorrectly claimed that modern experiments
cannot replicate these early observations.
In section~\ref{sec:Modern.expt}, 
we describe the reintroduction of this
phenomenon to the modern scientific community by a
secondary school student, Erasto Mpemba.
Mpemba's dramatic story cautions against
dismissing the observations of non-scientists, and
raises 
questions about the degree to which our theoretical
understanding can and should bias our acceptance and
interpretation of experiments.

In section~\ref{sec:Popper}, we see that that the 
Mpemba effect also provides a  good illustration of the
need to formulate a scientific hypothesis carefully,
and the need for theory in the design of an experiment.
We see that
the Mpemba effect is much harder
to study experimentally than it might appear at
first glance, and discuss some common problems
with lay experiments on the Mpemba effect.
Analysis of the experiments naturally brings up
Popper's thesis that a scientific
hypothesis must be falsifiable.

The discussions in
sections~\ref{sec:prehistory}-\ref{sec:Popper}
should be comprehensible regardless of why the Mpemba effect
occurs, and indeed, regardless of whether it 
occurs.
It is not until section~\ref{sec:theory} 
that we discuss possible theoretical 
mechanisms for
the effect, and readers uninterested in the history and
background can jump straight to this section. 
We explain why a common proof that the
Mpemba effect is impossible, is in fact flawed.
Multiple explanations have been proposed for the
effect. Evaporative cooling is one of the strongest
explanations, but the effects of convection, dissolved
gasses, and the surrounding environment, may all also be
important.
We discuss the results of modern experiments on the effect,
which are confusing, but for interesting reasons.

It has sometimes been reported that hot water pipes are more
likely to burst from freezing than adjacent cold water
pipes. Experiments on this phenomenon, which we 
discuss in section~\ref{sec:pipes}, have been more
conclusive than those on the Mpemba effect,
and target supercooling as the cause. 
The experiments on pipes are closely related to the
Mpemba effect, because they
provide a clear situation where the water can
``remember'' what has happened to it.
In section~\ref{sec:supercooling}, we look at the
possible importance of supercooling in the Mpemba effect,
ultimately concluding that its role is
unclear.
Finally, in section~\ref{sec:future} we discuss
possibilities for future experiments that could
be done by students.


\section{Experiments before the Scientific Revolution}
\label{sec:prehistory}

The Mpemba effect has long
been known in the Western world (although not by this
name until fairly recently).
Around 350 B.C., Aristotle wrote~\cite{Aristotle}

%
\smallskip
{\narrower
\noindent If water has been previously heated, this contributes 
to the rapidity with which it freezes: for it cools more
quickly. (Thus so many people when they want to cool water
quickly first stand it in the sun: and the inhabitants of
Pontus when they encamp on the ice to fish. . . pour hot
water on their rods because it freezes quicker, using the
ice like solder to fix their rods.) 
And water that condenses in the air in warm districts
and seasons gets hot quickly.

}
\smallskip

\noindent Aristotle used this observation in support of
antiperistasis, which is the 
``sudden increase in the intensity of a quality as a result of
being surrounded by its contrary quality, for instance,
the sudden heating of a warm body when surrounded by 
a cold.''~\cite{Marliani,antiperistasisProjectile}

While the idea of antiperistasis today sounds ridiculous,
with the hindsight of our modern understanding of heat,
energy, and temperature, it should be remembered
that Aristotle was working without these paradigms, and
indeed, without a thermometer. 
The fact that ice requires cold, yet hail comes in the
summer, rather than the winter, requires an
explanation---Aristotle's explanation was antiperistasis.
Later, a number of
medieval scientists used antiperistasis to explain the (apparent)
facts that bodies of water are colder in the summer,
and that human bodies are hotter in the 
winter~\cite{Marliani}.
While we can now explain these observations with
our modern theory of heat, the explanations are not
obvious.  The concept of temperature, and the
zeroth law of thermodynamics, are quite counterintuitive
to anyone who has touched metal and wood, outside
on a cold day.

In the 13th century, well before the Scientific Revolution,
Roger Bacon wrote {\it Opus Majus}, in which he argued
repeatedly for the importance of experiments in
science. He wrote~\cite{RogerBacon}

\smallskip
{\narrower
\noindent Moreover, it is generally believed that
hot water freezes more quickly than cold water in 
vessels, and the argument in support of this is advanced
that contrary is excited by contrary, just like enemies
meeting each other. But it is certain that cold water
freezes more quickly for any one who makes the experiment.
People attribute this to Aristotle in the second book of
Meteorologics; but he certainly does not
make this statement, but he does make one like it, by
which they have been deceived, namely, that if cold water
and hot water are poured on a cold place, as upon ice, the
hot water freezes more quickly, and this is true. But if
hot water and cold are placed in two vessels, the cold
will freeze more quickly. Therefore all things must be
verified by experience.

}
\smallskip

What is particularly interesting about this is that
Roger Bacon agrees that hot water can, {\it under some 
circumstances}, freeze faster than cold water, but
argues that specification of
the precise experimental conditions is important.
We will see that this is a crucial observation, 
equally important in discussions about modern
experiments on the Mpemba effect.

In the Middle Ages, debates raged over whether objects
could only be cooled by extrinsic sources, or whether some
objects might be able to
cool themselves.  In the middle of this debate,
around 1461, Giovanni Marliani
reported on experiments, described here by 
Clagett~\cite{Marliani2}:

\smallskip
{\narrower
\noindent . . .  To support his contention that heated water
freezes more rapidly [than cold], Marliani first points to 
a passage in Aristotle's {\it Meteorologica} affirming 
it. However,
[Marliani] does not depend on Aristotle's statement alone.
He claims that not only has he often tested its truth during
a very cold winter night, but that anyone may do so. You
take four ounces of boiling water and four ounces of
non-heated water and place them in similar containers. Then
the containers are exposed to the air on a cold winter's
morning at the same time. The result is that the boiling
water will freeze the faster.

}
\smallskip

The belief in the Mpemba effect continued strong into the 
17th century.
Francis Bacon and Descartes both
wrote extensive works on the scientific method, and
experiments, and both wrote about the Mpemba 
effect. In the {\it Novum Organum}, Francis Bacon
wrote~\cite{Bacon}

\smallskip
{\narrower
\noindent. . . water a little warmed is more easily frozen
than that which is quite cold. . .

}
\smallskip

\noindent And in 1637, Descartes wrote about 
this phenomenon in {\it Les Meteores}, a work which was published
attached to his more famous
{\it Discourse on Method}~\cite{Descartes}.
He emphasized the importance of
experiment, described how to analyze the 
density-dependence of water, and stated results
about the freezing times:

\smallskip
{\narrower
\noindent We can see this by experiment, if we fill a
beaker---or some other such container having a long,
straight neck---with hot water, and expose it to freezing
cold air; for the water level will go down visibly, little
by little, until the water reaches a certain level of 
coldness, after which it will gradually swell and rise,
until it is completely frozen.
Thus the same cold which will have condensed or shrunk 
it in the beginning will rarefy it afterwards.
{\it And we can also see by experiment that water which has been 
kept hot for a long time freezes faster than any other
sort}, because those of its parts which can least cease to
bend evaporate while it is being heated. (Emphasis added.)

}
\smallskip

A modern writer on Descartes has commented on the
italicized statement: ``This statement,
which the simplest of experiments could have refuted, was 
repeated with elaborate details in a letter to Mersenne,
and it emphasizes Descartes' readiness to rely on {\it $\grave{a}$ 
priori} conclusions''~\cite{DescartesModernComment}.
But this modern writer's position is contradicted by the letter 
to Mersenne, in which Descartes makes clear that he has, 
in fact, done this experiment. In this 1638 letter, Descrates 
wrote~\cite{DescartesLetter,DescartesLetterTranslated}:

\smallskip{
\narrower
\noindent I appreciate once again what you have written me
that my reputation is at stake in my response to Mr. Fermat,
in which I assure you that there is not one single word that
I would like to have changed. . . I dare to assure you that
there is nothing incorrect, because I did these experiments
myself, and particularly the one which you commented on of
the {\it hot} water that freezes more quickly than {\it 
cold}; where I said not {\it hot} and {\it cold}, but that
{\it water that one has held for a long time over the fire 
freezes more quickly than the other}; because in order to
correctly do this experiment, one must first have boiled
the water, then let it cool off, until it has the same
degree of coolness as that in a fountain, and having 
tested it with a thermometer, then
draw water from that fountain, and put the two waters in the
same quantity in same vases. But there are few people
who are capable of correctly doing these experiments, and often,
in doing them poorly, one finds the complete opposite of
what one should find. (Emphasis in original)

}
\smallskip

As with Roger Bacon's earlier experiment, we again see that
the details of the experiment are crucial. Descartes is
{\it not} measuring the time for the hot water to freeze, but is 
saying that when water has been heated, it is somehow changed
so that it cools more easily, even after being brought back to
room temperature. 
While the observation described is different than our modern
statement of the
Mpemba effect, it is similar in that
some sort of history-dependence (i.e. memory)
of the water is implied by the results.
Descartes's letter also indicates that both he, and others,
have done this experiment, and that the results are 
contradictory,
a problem that we will also see in
more modern experiments.

With the advent of the modern theory of heat, these earlier
observations were forgotten. Since these experiments {\it
appear} to contradict what we know about heat, it is 
natural to dismiss them as mistakes. 

Presentations in textbooks typically show the progress of
science as a simple, straight-line progression, with
experiments pointing in a straightforward and
unambiguous manner to new
scientific theories. But, as Kuhn has pointed out, the
development of scientific theories is not so simple~\cite{Kuhn}.
Most of the time, 
scientists are engaged
in what Kuhn calls ``normal science,'' 
during which research
``. . . is a strenuous and devoted attempt to force
nature into the conceptual boxes supplied by professional
education. . . .''~\cite{Kuhn2}
When scientists are working under a certain paradigm,
results that cannot be forced into the existing paradigm may
be ignored, as attention is focused on experiments that
appear more promising for advancing knowledge. To give but
one of Kuhn's examples~\cite{Kuhn3}:

\smallskip
{\narrower
\noindent
In the eighteenth century, for example, little attention was
paid to the experiments that measured electrical attention
with devices like the pan balance. Because they yielded
neither consistent nor simple results, they could not be
used to articulate the paradigm from which they derived.
Therefore, they remained mere facts, unrelated and
unrelatable to the continuing progress of electrical
research. Only in retrospect, posessed of a subsequent
paradigm, can we see what characteristics of electrical
phenomena they display.

}
\smallskip

\noindent Whatever one's opinion on Kuhn's more
controversial theories, it is clear that
scientists' theoretical views
influence 
what experiments they choose to do, to trust and to think about.

The Mpemba effect illustrates the
points raised by Kuhn. 
In modern times, because the Mpemba effect {\it appears}
to contradict modern theories of heat, scientists
are much more skeptical, or even dismissive, of
experiments that observe the Mpemba effect.
Furthermore, like the eighteenth century experiments with
pan balances, experiments on the Mpemba effect,
for reasons we will discuss,
yield ``neither consistent nor simple results.'' 
Thus, the Mpemba effect,
while interesting, is a factual curiosity,
and not fundamental to 
our modern understanding of heat.

In {\it The Structure of
Scientific Revolutions},
Kuhn briefly mentions these experiments by
Marliani and Bacon~\cite{Kuhn4}:

\smallskip
{\narrower
\noindent. . . the natural histories often juxtapose
[correct] descriptions. . . with others, say, 
heating by antiperistasis (or
by cooling), that we are now quite unable to confirm.

}
\smallskip

\noindent Kuhn does not cite any experimental evidence
that we are unable to confirm these older results,
and one suspects that he is simply assuming this 
point~\cite{KuhnAssumptionNote}.
In fact, as we will see, at the time that Kuhn wrote this,
there were multiple 
papers confirming the existence of the Mpemba effect.
Kuhn thus unintentionally, and ironically,
demonstrates how our
theoretical expectations can color our
experimental beliefs.


\section{Erasto B. Mpemba and 20th century knowledge}
\label{sec:Modern.expt}

This strange phenomenon was reintroduced to the 
modern scientific
community by Erasto Mpemba, a secondary school student 
in Tanzania, in 1963. Mpemba told his story
in {\it Physics Education}~\cite{Mpemba.story}. 
Mpemba and his fellow students were
making ice cream, which used a mixture that included
boiled milk.
Because excessively hot objects could damage
the refrigerator, they were supposed to let their mixture
cool before putting it in the refrigerator. However,
space in the refrigerator was scarce, and when another
student put his mixture in without boiling the milk,
Mpemba decided to put his hot mixture in, without
waiting for it to cool. Later, Mpemba found that his hot
mixture froze first.

Mpemba asked his teacher for an explanation, but his teacher
said Mpemba must have been confused. When
the teacher later covered Newton's law of cooling,
Mpemba persisted in his questioning. The teacher responded
that ``Well, all I can say is that is Mpemba physics and not
the universal physics,'' and from then on the teacher
and the class would mock his mistakes by saying 
``That is Mpemba's mathematics,'' or ``That is 
Mpemba's physics.'' 

Mpemba ran more systematic experiments, both
with water and milk, and continued to get 
similar results. When Dr. Osborne, a professor at a 
nearby university, visited Mpemba's school, Mpemba asked 
him why water at $100^O C$ froze faster than water
at $35^O C$.  
Upon returning to his university, Dr. Osborne asked a 
technician to test Mpemba's question. When Dr. Osborne
later asked for the results,
``The technician reported that the water that
started hot did indeed freeze first and added in a moment
of unscientific enthusiasm 
{\it `But we'll keep on repeating the
experiment until we get the right result'} ''
(emphasis added)~\cite{Mpemba.story.quote}.
But future experiments gave similar results, and 
Mpemba and Osborne later published their
results~\cite{Mpemba.story}. In the same
year, Dr. Kell of Canada independently 
reported the phenomenon, along with a theoretical
explanation that we will consider later~\cite{Kell}.

Subsequent discussions in journals
showed that the effect was
already known and believed by non-scientists in
diverse regions of the world.
Kell stated that it was widely believed in Canada, and
that 

\smallskip
{\narrower
\noindent Some will say that a car should not be washed with hot
water because the water will freeze on it more quickly than
cold water will, or that a skating rink should be flooded
with hot water because it will freeze more quickly.

}
\smallskip

\noindent Mpemba reported that after his initial experience,
he found that ice cream makers in Tanga City, Tanzania, 
used hot liquids to make the ice cream faster. 
British letters to the {\it New Scientist} 
reveal a 
wealth of lay observations. One writer stated that it 
was well-known that in the winter, hot water
pipes were more likely to freeze than cold water
pipes~\cite{pipesLetter} (an issue we consider in
section~\ref{sec:pipes}). Another stated that the 
phenomenon was well-known in the food-freezing
industry~\cite{food}. A number of letters reported
that their (non-scientist) friends and family had
known of this effect even in the
1920's~\cite{old.1,old.2,old.3}.
One writer was a science teacher, whose story closely
parallels Mpemba's. The teacher 
describes his experiences with a student who asked him
why hot water froze faster
than cold\cite{Parallel.Story}

\smallskip
{\narrower
. . . I told him that it was most unlikely, but he replied
that he had seen it happen when his mother threw out her
washing water onto the path. I explained to him that the
particles in the hot liquid would be escaping more
readily to the atmosphere due to evaporation and that this
would leave a thinner layer of liquid to freeze than in the
colder one where the particles would be escaping more
slowly. He was not, however, convinced, and a few days later
he reported that he had placed two cans, one containing hot
water, and the other cold water, outside and that the hot
one was still the first to freeze. Finally, he challenged me
to explain that one, if I could. Still in no doubt I
criticized his experiment and suggested he attempt a more
accurate one under laboratory conditions.

He obtained two identical specimen jars and placed hot
water (approx $50^O$ C) in one and colder water
(approx $20^O$ C) in the other; both tops were left off and
they were placed in the freezing compartment of a
refrigerator. To my disbelief, and his delight, the hotter
one was indeed the first to form a layer of ice on the
surface.

He was, to say the least, no longer quite so impressed 
by my capabilities as a Science Teacher.

}
\smallskip

The skepticism with which scientists react
to the Mpemba effect is quite common. In this author's
experience, scientists are much
more likely to react with disbelief than laypersons; 
this is not surprising, since scientists know enough to 
``know that this is impossible.'' 

Mpemba's story provides a dramatic parable against
dismissing the observations of non-scientists.
But his story is particularly interesting because it
is more than just a story of bad, close-minded,
scientists. There is, of course, no excuse for the
Tanzanian teacher's mockery of Mpemba. But is it ``unscientific'' 
for scientists to, upon hearing of the Mpemba effect,
immediately suspect errors in the experiment?
Kuhn emphasizes that 
with the operation of ``normal science,''
scientists interpret
experiments in light of the reigning paradigm---i.e.
with their preexisting theoretical
understanding~\cite{Kuhn}. 

What is interesting about the Mpemba effect
is that unlike the examples
commonly given in science textbooks, where theory and experiment
march hand-in-hand, always leading to further progress,
here theory (rightly or wrongly)
prevents acceptance of experiment. 
We are certainly not arguing that
the reaction of the scientific community to 
surprising experimental results is arbitrary,
or necessarily hostile. Our point is simply that
the reaction to an experiment 
depends significantly 
on how well that experiment matches
accepted theoretical preconceptions. 
Because experimental claims can be in error,
scientists do not just accept all published claims.
While few scientists would
find this claim controversial, it is quite different
than the impression one gets from science textbooks,
and from what appears in certain positivistic
views of science. The Mpemba effect provides
a lovely case for considering these issues, because
while it provokes skepticism, it has been observed
in multiple experiments; yet, in support of the skeptical
position,  we will see that
the experimental results are not entirely
consistent, and that the theoretical situation is still 
unsettled.


\section{What is the question, and is it scientific?}
\label{sec:Popper}

To analyze the Mpemba effect,
we first need to 
precisely state what we are trying to test.
At first sight, the question is quite
simple: ``Does hot water freeze faster than cold?''
However, a little thought shows that this formulation is not
adequate. Clearly, a small drop of hot water can
freeze faster than a cold ocean. Hot water in a freezer will
freeze faster than cold water on a warm day (as the latter
will not freeze at all). These examples are silly, 
but illustrate the need to state the question clearly.

A better second attempt
at stating the problem might be ``Given two
bodies of water, which are identical in 
all parameters (mass, shape, surroundings, etc. . .),
except that
one is initially at a higher uniform temperature than
the other, the hotter water will freeze first.''
But further thought shows that
this cannot be correct. 
If the initially hotter water
is at $99.9999^O C$, and the initially colder water is 
at $0.0001^OC$, then the initially cold water is just
seconds away from freezing, and it is clear
that the hot water cannot possibly
overtake it.
Furthermore, there is clearly no reason to expect the
Mpemba effect to occur for {\it all} possible initial
parameters.

So a better phrasing might be
``There exists a set of initial parameters,
and a pair of temperatures, such that given two bodies
of water identical in these parameters, and differing
only in their initial uniform
temperatures, the hot one will freeze
sooner.'' This is a much better statement of the question,
although we will see later that
deficiencies remain.

Once the Mpemba effect is properly stated, it is clear that
we are only looking for {\it some}
set of parameters, such that if we plot
freezing time versus initial temperature, there
is {\it some} range of the graph that is downwards
sloping---not that 
there
exist initial parameters such that the graph is
downwards-sloping all the way from $0^O C$ to
$100^O C$, or that
the graph has downwards-sloping parts
for any set of initial parameters.
This is logically necessary for the problem
to be at all reasonable, but 
this point is not always appreciated
in popular discussions. 
Consider the  discussion of the Mpemba effect in Ann
Landers' column, as described in
{\it Myth-Informed}~\cite{Myth.book}

\smallskip
{\narrower
\noindent [Does] warm water [freeze] faster than cold
water. . .
The redoubtable Ann Landers got into an ongoing row with
readers in 1983 when she addressed this question, as well
as the related cosmic issue of whether cold water boils
faster than hot water. She ``went to the top'' and
consulted Dr. Jermore Weisner, chancellor of the
Massacusetts Institute of Technology, who kicked the
problem over to the MIT dean of science, Dr. John
W. Deutch. Landers never recorded what Deutch thought of
being given such a problem by an advice column, but the
eminent scientist reported ``Neither statement was true.''
Whereupon ``Self-Reliant in Riverdale''. . .
upbraided her for using ``argument
by authority'' rather than doing her own experiment.
``Self-reliant'' said she reached the same conclusion as
Deutch by using a pan of hot water, a thermometer, a
stove, a refrigerator, and a watch with a second hand.

}
\smallskip

In his popular column, {\it The Straight Dope}, Cecil
Adams also discussed the Mpemba effect~\cite{StraightDope}:

\smallskip
{\narrower
\noindent . . . I carefully measured a whole passel of
water into the Straight Dope tea kettle and boiled it for
about five minutes. This was so I could compare the freezing
rate of boiled $H_2O$ with that of regular hot water from
the tap. (Somehow I had the idea that water that had been
boiled would freeze faster.)

Finally I put equal quantities of each type into trays in
the freezer, checked the temp (125 degrees Fahrenheit all
around), and sat back to wait, timing the process with my
brand new Swatch watch, whose precision and smart styling
have made it the number one choice of scientists the world
over.

I subsequently did the same with two trays of cold water,
which had been chilled down to a starting temperature of 38
degrees.

The results? The cold water froze about 10 or 15 minutes
faster than the hot water, and there was no detectable
difference between the boiled water and the other kind.
Another old wives' tale thus emphatically bites the dust.
Science marches on.

}
\smallskip

These discussions fail to appreciate that
a single test
cannot show that the effect never occurs for any
parameters and initial temperatures.

Further consideration of this point brings up another
issue. Logically,
our statement about the Mpemba effect can never
be proven false. Regardless of how many experiments
Cecil Adams runs, a true believer in the Mpemba effect
can always claim that the effect occurs for other
sets of initial parameters, that differ slightly
from the ones that Cecil Adams used. Popper has
famously argued that the hallmark of a scientific
hypothesis is that it be falsifiable~\cite{Popper}.
That is, not that it can be proven true, but that it 
can be proven false.
Is our most recent statement of the Mpemba effect
thus unscientific? 

It is not unusual that a scientific phenomenon,
strictly speaking, cannot be proven impossible,
because the parameter space in which it might
occur is, in principle, infinite. However, if 
we search a 
``representative sample'' of 
the parameter space over which the phenomenon might
be thought to occur, and the phenomenon
is not seen, that would be fairly convincing evidence
against it. 


We thus need a list of experimental parameters that we might
vary when studying the 
Mpemba effect.
The list is rather long (infinite).
On a first page, we might list the mass of the
water, the shape of the container, the surrounding
environment of the refrigerator, and the gas content of the
water. Note that several items on this list are not
single parameters---characterizing the shape
of the container requires multiple parameters.
We may also
want to include boolean parameters, such
as whether the container has a lid.
On a second page, we might list
the color of the container, 
the electrical conductivity of the walls of the
refrigerator, the gender of the experimenter, 
etc. . . If we simply list all the parameters
we might think of, the list is infinite, and we
are at a complete loss as to how to proceed with an
experiment. Without a theoretical framework in which
to design and conduct the experiment, we are reduced to
randomly collecting facts, such as the color of the
container, and the situation is impossible. 
However, we have strong theoretical reasons for
ignoring all the parameters on the second page, and
the experimenter can safely ignore the color of the
container, neither varying, nor recording, its
color.

Unfortunately, as we will see in next section,
all the items of the first page can plausibly be 
considered important for the Mpemba effect.
Furthermore, their effects are not 
independent of one another.
But an experimenter 
can hardly be expected to establish
a vast multidimensional array of containers
of different dimensions and shapes, independently 
varying masses, gas contents, and 
refrigeration methods. 

We are, of course, not claiming that scientific investigation
of the Mpemba effect is impossible. But a common 
response, upon first hearing of the Mpemba effect,
is that it should be a straightforward matter to 
study experimentally. 
But for even this (deceptively) simple
problem, productive experimental design requires at least
some theoretical understanding of why the effect might
occur---otherwise we will not know whether we should 
be paying attention to, for example, the gas content of the water.
We will see in the next section that because the time
to freezing is sensitive to all the 
``first page'' parameters,
the experimental results become very confusing.

Our statement of the Mpemba effect now reads
``There exists a set of initial parameters
(water mass, gas content of water,
container shape and type, and refrigeration method),
and a pair of temperatures, such that given two bodies
of water identical in these parameters, and differing
only in their temperatures, the hot one will freeze
sooner.'' One final difficulty that must be 
considered is how to
define the time of freezing---are we considering 
it frozen when ice crystals first appear, or only
when the entire body of water is frozen? Or,
to simplify the experiment,
we might just measure the
time until 
some specified part of the water reaches $0^O$ C.
This might seem like a minor issue,
but we will see in 
section~\ref{sec:supercooling}, on supercooling,
that it is potentially crucial.


\section{How could the Mpemba effect occur?}
\label{sec:theory}

We have deliberately avoided discussing the 
theoretical explanations for the Mpemba effect
until now. We have done this both because the
historical reaction to the Mpemba
effect is only comprehensible in light of the effect's
{\it apparent} inconsistency with modern
conceptions of heat, and to emphasize the 
need for an experimental framework when designing experiments 
on the effect.  Here, we discuss some proposed mechanisms
for the Mpemba effect, but will not attempt to analyze
their relative plausibility in depth.

To see how the effect might occur, it is useful to
carefully 
think about why the effect appears
impossible. The careful reader may already have come up
with a proof of impossibility that goes something like this:

\smallskip
{\narrower
\noindent Suppose that the initial temperatures for the hot
and cold water are $70^O C$ and $30^O C$. 
Then the $70^O C$ must first cool to $30^O C$, 
after which it must do everything the $30^O C$
water must do. Since the $70^O C$ water has to do
everything the $30^O C$ water must do, plus a little
more, it must take longer to freeze.

}
\smallskip

\noindent A good way of systematically analyzing the
Mpemba effect is to think about why this proof is
wrong. 

The problem with this proof is that it implicitly
assumes that the water is completely characterized
by a single parameter---the temperature.
We need to think of a parameter
that might change during the course of the experiment;
then, the $70^O C$ water cooled to $30^O C$
will not be the same as the water initially
at $30^O C$. 

One possible parameter is the mass of the water.
Both bodies of water initially
have the same mass. But if the initially hotter water
loses mass to evaporation, then the $70^O C$ water
cooled to $30^O C$ will, having less mass, be easier
to freeze (i.e. less energy will need to be removed to
cool and freeze it). This is one of the strongest
theoretical explanations for the Mpemba effect.
Kell numerically integrated the heat loss equations,
assuming that the cooling was by evaporation alone,
and that the mass lost to evaporation never
recondensed----he found that 
with these assumptions, there were initial temperatures
where the hot water would freeze faster than the cold 
water~\cite{Kell}. But this does not
prove that evaporative cooling is the {\it only} factor
behind the Mpemba effect.
A number of experimenters have
claimed that the amount of mass they lost to evaporation was
insufficient to explain their 
results~\cite{Osborne2,Osborne3,Freeman}.
And Wojciechowski et. al. observed
the Mpemba efect in a {\it closed} container, which
further suggests that evaporative cooling is not
the sole cause of the effect~\cite{Wojcie}.

A more complex ``parameter'' is the temperature distribution
of the water. 
As the water cools it will develop convection
currents, and the temperature will become non-uniform. 
This also defeats the impossibility proof 
above, since the water is no longer characterized by a
single number. Analysis of the situation is now
quite complex, since we are no longer considering a single
parameter, but a scalar function, and computational fluid
dynamics is notoriously difficult.
Nevertheless, at least one general
point can be made. For temperatures above $4^O C$, hot water 
is less dense than cold water, and will thus rise to the
top. So we can generally expect that when the $70^O C$
water has cooled to an average temperature of $30^O C$,
the top of the water will be hotter than $30^O C$, and the bottom
of the water will be below $30^O C$. If the water 
primarily cools at its surface, the
non-uniform distribution with an average temperature
of $30^O C$ will thus lose heat faster than 
uniformly $30^O C$ water. 
Consistent with this, Deeson found that gentle
stirring substantially raised to time to
freezing~\cite{Deeson}.
Convection could work in concert with other factors, such as
evaporative cooling.
Convection currents are sensitive to the
shape and dimensions of the container,
so this explanation may play very different roles
for different container shapes.

This brings up the question of what ``parameters''
should be associated with the surrounding air.
Modeling the cooling process should take into account the
convection currents of the air, which will depend on
the shape of the refrigerator. 
Firth's studies of the Mpemba effect found that this 
was an important factor~\cite{Firth}:

\smallskip
{\narrower
\noindent What these experiments have shown, however, is that
it is not just the beaker or the initial temperature of
its contents that are important, but that its environment
probably influences the cooling rates to a greater extent
than any aspect of the beaker itself.

}
\smallskip

On a related note,
one letter writer to {\it The New Scientist}
suggested that the Mpemba effect could be explained if
the containers of water were sitting on layers of frost.
He argued that
the frost conducts heat poorly, and the hot water
causes the layer of frost to melt, establishing better
thermal conduct with the refrigerator floor~\cite{frost}.
This may explain some everyday observations of the
effect, but the
experiments published in refereed
journals generally used containers on thermal insulators.

Another possibility is that the hot and cold water, while
they appear identical to the naked eye, differ in their
gas content. Hot water can hold less
dissolved gas than cold water, and the gas content
affects the properties of the water.
Mpemba and Osborne's original experiments were
done with recently boiled water, to remove 
dissolved air~\cite{Mpemba.story}, as were the
experiments by Walker~\cite{Walker}.
These experiments suggest that dissolved gasses are not
necessary to the Mpemba effect. (Although 
under typical conditions, the degassed
water will slowly regain dissolved gasses from the
atmosphere, confusing matters.)
On the
other hand, Freeman 
only observed the Mpemba
effect when carbon dioxide was 
dissolved in the water~\cite{Freeman}.
Similarly, Wojciechowski et. al.
only saw the effect for
non-degassed water~\cite{Wojcie}.
A number of explanations
of have been proposed for how the amount of dissolved
gas could affect the properties of the water, and thus
cause the Mpemba effect, although they remain largely
speculative. One of the few quantitative findings
was by Wojciechowski et. al., who
reported that for water
saturated with carbon dioxide, the enthalpy of 
freezing was smaller for the initially warmer
water (but that preheating was irrelevant to the
enthalphy if dissolved gasses were absent)~\cite{Wojcie}. 
We will return to the effects of dissolved
gases, and other impurities, when we discuss 
supercooling.

All the factors discussed are at least potentially important
in explaining the Mpemba effect. What makes the situation
particularly difficult to analyze is that the factors are
not independent of each other---for example, rates of
evaporative
cooling will depend on the shape of the container. 
The experimental results described here
do not all point to a single, clear, 
conclusion, and the reader who further inspects the
original papers will 
find more facts, but will
not find the overall picture much
clearer~\cite{Mpemba.story,Freeman,Wojcie,Walker,Firth,Auerbach,Deeson}. 

Because there are so many factors that can be varied, and
the results of the experiments can depend sensitively on any
of these factors, experimental results are varied and 
difficult to organize into a consistent picture. (Recall
Kuhn's statement about 18th century pan balance 
experiments, quoted in section~\ref{sec:prehistory}.) It is
thus unclear which of these explanations is 
{\it the} explanation, or indeed, 
whether it is appropriate to look for a single
explanatory factor, isolated from the other factors. 
As Firth wrote~\cite{Firth},

\smallskip
{\narrower
\noindent There is a wealth
of experimental variation in the problem so that any
laboratory undertaking such investigations is guaranteed
different results from all others.

}
\smallskip

\begin{figure*}[tb]
\epsfig{figure=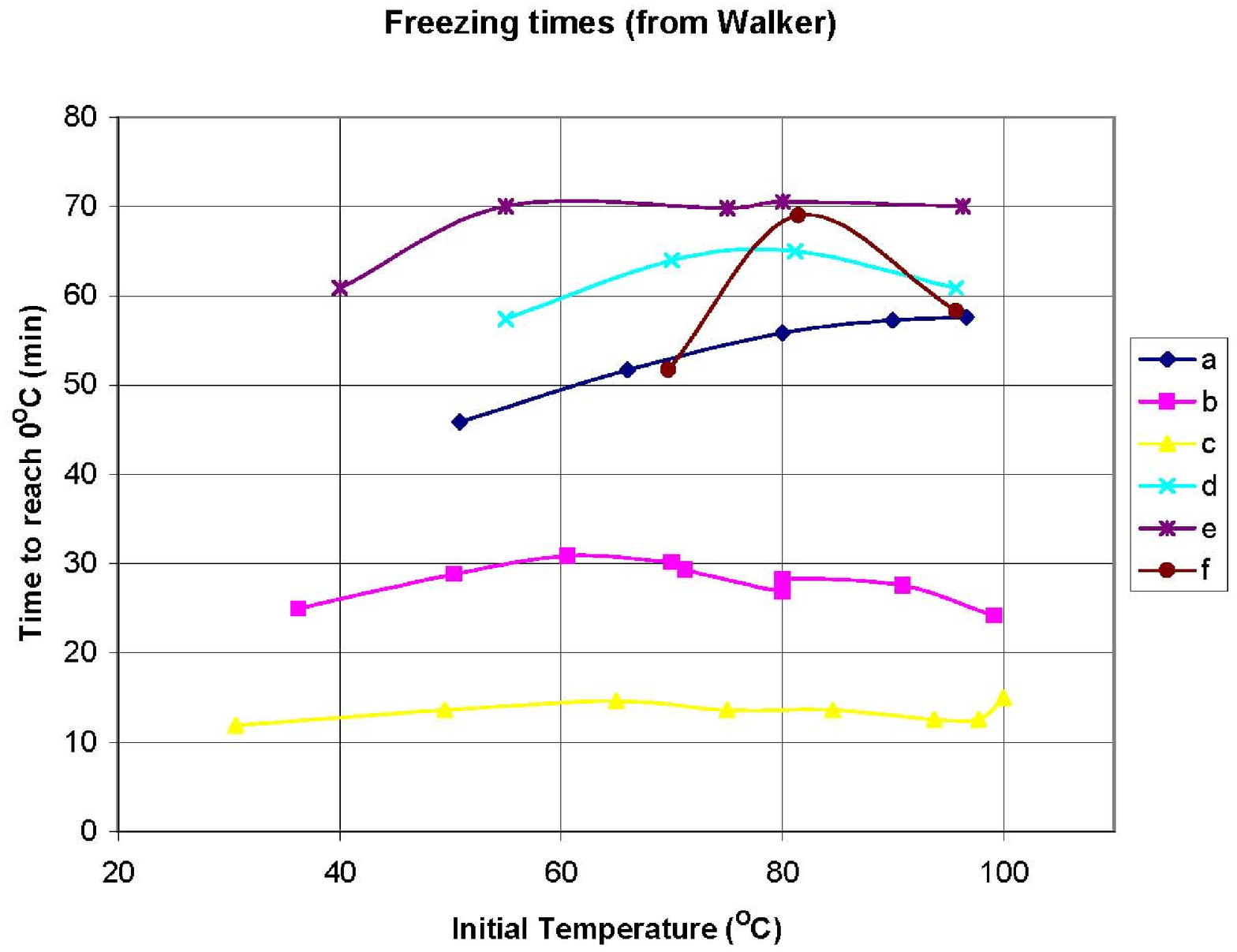,width=5.8in}
\caption{Dependence of time of freezing on initial
temperature, for various experimental conditions:
(a) 50 ml in small beaker (b) 50 ml in large beaker
(c) 50 ml in large beaker in frost-free freezer
(d) 100 ml in large beaker, thermocouple near bottom
(e) 100 ml in large beaker, covered with plastic wrap,
thermocouple near bottom
(f) 100 ml in large beaker, thermocouple near top.
Graph produced from data
obtained by Walker~\cite{Walker}.}
\label{fig:Walker}
\end{figure*}

In figure~\ref{fig:Walker} we see experimental results
by Walker, showing the dependence of the time of
freezing on the initial temperature,
under various initial conditions. We see that some
graphs show a strong Mpemba effect, some only show a weak
one, and some show no Mpemba effect at all.
This indicates that
the cooling is indeed sensitive to a number of parameters.
Furthermore, Walker reported that while his results were mostly
repeatable, he ``still obtained strange large deviations on
some of the results.''~\cite{Walker}.

\section{Supercooling and Bursting water pipes}
\label{sec:pipes}

It has often 
been said that hot water pipes burst from freezing more 
often than adjacent cold water 
pipes~\cite{pipesLetter,Brown,Dorsey,Gilpin}.
This is different from the Mpemba effect, but it is
similar to it, in requiring the water to have
a ``memory.'' The mechanism behind the bursting water
pipes is better understood than that behind the Mpemba
effect.

The water pipe claim was first investigated by F. C. Brown, in 
1916~\cite{Brown}.
He confirmed the claim by taking 100 glass test tubes, and
filling 50 with tap water, and 50 with tap water that had been
boiled. {\it After allowing all water to 
first reach room temperature},
he placed them outside in subzero temperatures. He found that 44 of the 
tubes with the boiled water burst, while only 4 of the tubes with
non-boiled water first. Since all water was at the same temperature
when placed outside
(as with Descartes's experiment),
and he was looking at the occurence of bursting,
rather than the time until freezing, these are not
tests of the
Mpemba effect. 

Freezing water will generally
supercool. Supercooling to $-4^OC$ to $-6^OC$ is common, and much
greater supercooling can occur for small
samples~\cite{Dorsey,Knight}. 
Once freezing starts, the ice-water mixture must go to
$0^O C$.
So when freezing begins, a finite
fraction of the water must lose energy, and turn to ice, 
transferring heat to the remaining subzero water, whose
temperature will rise to $0^O C$. Thus, the more supercooling
occurs (i.e. the lower the temperature at which 
freezing begins), the larger the volume fraction of water that 
will freeze initially (i.e. the larger the fraction of 
$H_2 O$ molecules that will be in ice structures).
Also, a certain volume fraction of ice will {\it not}
always correspond to the same 
volume of region spanned by the ice.
The ice will form a dendritic structure,
interspersed with liquid. If more supercooling
has occurred, more of the water will have reached subzero
temperatures, and the dendritic structure will span
a larger region, for a fixed volume fraction of ice.

Brown observed that the water
that had been boiled first, later supercooled, while the nonboiled
water did not, and argued that this difference was responsible for
the difference in bursting behavior~\cite{Brown}.
He argued as follows:
If water in a pipe freezes near $0^O C$, only a small amount of 
ice will be formed initially. This ice will be localized to the
coldest regions, at the sides of the pipes, leaving a hole
in the center. With further cooling, the 
hole will shrink, but until all the water has frozen, liquid
water will still be able to flow through the hole.
Furthermore, the water flow can
break away the ice on the sides,
relieving pressure.
On the other hand, if the water 
supercools significantly before freezing, 
a larger structure of dendritic 
ice will form throughout the pipe---while only a finite
fraction of the water will be turned to ice, this 
dendritic ice will span more of the pipe, possibly
leaving no hole, thus
blocking the flow of water, and resulting in a burst pipe.
Consistent with this explanation, Brown found that the
ice rose higher from expansion in the tubes with
nonboiled water,
indicating the existence of a
central region in which the water was mobile, and the
ice able to push up.

\begin{figure}[tb]
\epsfig{figure=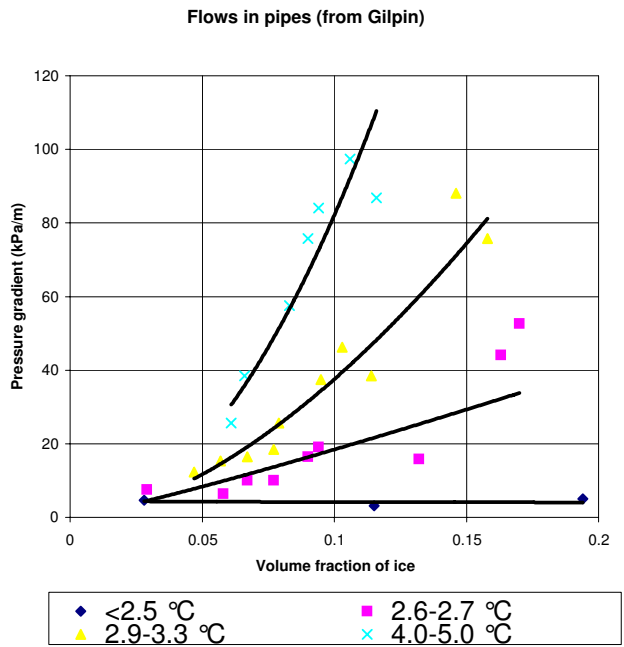,width=3.3in}
\caption{Effect of nucleation temperature on
pressure gradient required to start flow through pipe.
Each of the four sets of data points are for different
amounts of supercooling.
Graph produced from data obtained by
Gilpin~\cite{Gilpin}.
Curves are power law fits, and are guides to the eye only (no 
theoretical justification.)}
\label{fig:Gilpin}
\end{figure}

In 1977, Gilpin performed quantitative
experiments that
confirmed Brown's
qualitative explanation~\cite{Gilpin}. 
Gilpin exposed pipes to subzero
temperatures, and
measured the pressure gradient necessary to
induce the flow of water, at various times after the 
ice had formed. He found 
that the more
supercooling, the greater the pressure gradient
needed, {\it for the same amount of total ice formed}
(figure~\ref{fig:Gilpin}).
He concluded, as Brown had, that with more supercooling, the
ice formed would be more likely to form a dendritic
structure,
spanning the pipe, and causing blockage.
He also confirmed this picture with photographs of the 
cross-section of the pipe during the freezing process.

These experiments convincingly demonstrated
that greater supercooling would lead to burst pipes,
leading to the question of why initially hot water would supercool
more than initially cold water.
Brown argued it was because 
boiled water had less dissolved gas, and that the
dissolved gas prevented supercooling~\cite{Brown}.
However, Dorsey, who carried out an extensive series
of experiments 
on the factors affecting supercooling, 
over the course of ten years, disputed
this~\cite{Dorsey,KnightDorsey}. Dorsey found that
dissolved gasses were not a significant barrier to
supercooling; he also pointed out that, unlike boiled
water, the water in hot water pipes does
contain significant amounts of dissolved gas. Dorsey
agreed that heated water supercooled more, and that
this would result in burst pipes, but argued
that the greater supercooling occurred because heating
served to deactivate or expel ``motes'' (what we would
today call nucleation sites)~\cite{Dorsey}.
Both explanations agree that some sorts of nucleation sites
are deactivated by heating the water. Gilpin not only
confirmed that hot tap water supercooled more than cold tap
water, but that tap water left in an open container
supercooled least of all---this can be explained by the fact
that water in an open container will absorb impurities from
the air, and these impurities can then act as nucleation 
sites~\cite{Gilpin}.

Again, the phenomenon here is not the same as the Mpemba
effect, but is related, in that it explains how the water can
remember its history. However, note that the results here
are in the wrong direction to explain the Mpemba effect.
If initially hot water supercools more, then to freeze it has
to reach an even lower temperature than initially cold water,
which will lengthen the time that it takes to freeze.


\section{Supercooling and the Mpemba effect}
\label{sec:supercooling}

Consideration of supercooling greatly complicates
the Mpemba effect, and it is not clear how, or whether, it
helps to explain it. We first
need to
decide precisely how we measure 
the ``time to freezing.''
If we wait until
the first appearance of ice,
then the experimental situation is
complicated by the randomness of supercooling, and multiple
trials will be needed to get precise average times to
freezing. For simplicity, many experiments have studied the
time for some specified location in the water to reach
$0^O$ C~\cite{Firth,Walker,Freeman}.
Supercooling is irrelevant for these experiments, if
the specified location is at or near the place that first
reaches $0^O$ C.

Auerbach  considered 
the relevance of
supercooling to the effect~\cite{Auerbach}. He found
that initially hot water would
supercool less (measuring the time to the first appearance 
of ice crystals) than initially cold water.
Auerbach did not determine why this happened, but
pointed out that the initially hotter water should have greater
temperature shear, and that this shear is known to trigger
crystallization~\cite{TempShear}.
However, Auerbach's finding that heated water
supercools less than nonheated water is directly
opposite to the findings of Brown and Dorsey, described
in the previous section.
Furthermore, Auerbach had a relatively small number of
trials, so the significance of his results is unclear.

While Auerbach's result is in the correct
direction to explain the Mpemba effect (since if the
hot water supercools less, this will help it freeze
sooner),
he actually found that the initially hot water actually
took longer to freeze on average (due to the
greater time the hot water took to reach 
$0^O C$)~\cite{Auerbach}.

Given this, it is unclear whether Auerbach's experiments
can be described as actually observing the Mpemba effect.
Due to the random fluctuations in the
times till freezing, Auerbach found that the hot
water might freeze first by chance.
He found that when the ambient temperature was
$-5^O C >T_a>-8^O C$, the probability for a randomly
chosen container of 
initially hotter water to freeze before a randomly
chosen container of
initially colder water was 53\%~\cite{Auerbach}. 
For ambient
temperatures of
$-8^O C>T_a>-11^O C$, the probability was 24\%.
If the hot water, on average,
takes longer to freeze, but only happens to freezes first 
in some specific samples due to random fluctuations,
it is not clear that
this should be called a Mpemba
effect.
(Given Auerbach's small number of samples, 
53\% is not signficantly greater than 50\%.)
While the experiments on pipes show that supercooling can
induce significant memory effects, the
role of supercooling in the Mpemba effect 
remains uncertain.


\section{Future prospects}
\label{sec:future}

It is clear that evaporative cooling can play an important
role in the Mpemba effect, and
that the history of the water can affect the amount of
supercooling. But beyond that, the experiments together
paint a very muddled picture. More experiments are needed
to solve this 2000+ year-old puzzle. Despite the
theoretical complexity of the Mpemba effect, the experiments
needed to probe it can be done at the K-12 and undergraduate
level. Indeed, simple experiments on the Mpemba effect are a 
common science fair project.

A fair amount of thought needs to go into the experimental
design
before even the first data point is taken. Walker
discusses the basic set-up, and while you can
come up with your own design, reading Walker's article
is a good way to appreciate some of the hidden complexities
in the experiment~\cite{Walker}. For example, Walker
points out that the container should be heated along with
the water, for if hot water is poured into a cold
container, the sudden change in the water's temperature
causes problems. To make sure that all samples of water have the
same mass, masses need to be measured
after heating, rather than before, as a fair amount
of mass is lost during the heating process. 
The precise environment surrounding the container is
important, and it can make a difference whether the 
water is in a middle of an empty freezer, or jammed 
between a frozen pizza and a frost-covered
bucket of ice cream. The temperature can be
read with a common mercury thermometer, but a device that can more
quickly and accurately register changes in temperature is better, if 
available. The reader is directed to Walker's article for
further discussion of potential problems and
solutions~\cite{Walker}.

A series of trials will produce a graph of freezing
time vs. initial temperature, of the sort shown
in figure~\ref{fig:Walker}.
A single curve may or may not show a Mpemba effect, but is not, 
on its own, particularly useful for probing the cause(s) of this
phenomenon. To see how the Mpemba effect depends on the various
parameters of interest (initial mass, gas content, container
shape and type, etc. . .) several curves need to be produced,
under different parameter settings. If
you had infinite resources and time, you could vary all the
initial parameters, producing a giant multidimensional
matrix of freezing times.
In practice, you will have to decide
which factors you are most interested in, and design your
experiment accordingly. 

You can decide on your own what to vary to produce your series of curves. 
I give some suggestions below, but the number of reasonable
experimental designs is essentially unlimited.

Kell claimed that because the Mpemba effect relies on
surface cooling, it is more likely to be observed
in wooden pails than in metal ones, since in metal pails a
fair amount of heat is lost through the sides. This
claim can be tested by producing a series of curves of
freezing time vs. initial temperature, for containers with
differing degrees of thermal insulation on the sides.
This would demonstrate how the Mpemba effect is affected
as the relative importance of evaporation to the cooling process
is varied.
Alternatively, varying the
height of the water, while keeping the base fixed, provides
another way of varying the importance of evaporation.

The importance of evaporation can be largely eliminated
by putting the water in a closed container, or by
putting a layer of oil over the surface of the water.
Such experiments would be extremely useful in assessing
claims that evaporation is {\it the} cause of the Mpemba
effect. As already discussed, several authors have claimed
that evaporation is insufficient to explain their results. But I 
know of only one published paper studying the Mpemba effect in
a closed container~\cite{Wojcie}, and a single experiment
can always be in error. If you run a series of
experiments in closed containers, and regularly fail to find 
a Mpemba effect, that would provide good support for the claims
that evaporation is the primary cause of the effect. On the other
hand, observations of the Mpemba effect in closed containers 
would show
that it can occur without evaporation.

Rather than looking at freezing time versus initial temperature,
you could investigate supercooling. Simply reproducing earlier
experiments would be valuable in resolving the discrepancy
between the recent results of Auerbach, and the older
results of Brown and Dorsey~\cite{Brown,Dorsey,Auerbach}.
The number of runs in Auerbach's experiment was too limited
for firm conclusions to be drawn, so a repetition of his
experiment with a greater number of runs would be useful.
A modern repetition of Descartes' experiment would also be
interesting.

More systematic studies of how the 
history of the water affects its properties would be
helpful.
For example, if you find that pre-boiled tap water has 
different properties than water straight out of the tap,
you could investigate why it differs (dissolved oxygen?
impurities?), by looking at how long it takes pre-boiled 
water's properties to return to those of tap water's, and 
under what conditions.  

You should try rerunning the experiments several times with 
exactly the same parameters, to get an idea of how big the error
bars are. The size of the error bars is crucial, since if a
graph of freezing time vs. initial temperature shows only a 
weak local maximum, it will be unclear if this is a true Mpemba
effect, or just the result of fluctuations. And, as with any
experiment, you want to make sure that your results are
repeatable. This greatly increases the time required for the experiment,
and limits the amount you can vary the parameters, but it's better
to have a small amount of reliable data than a large amount of
unreliable data!

What will constitute experimental success? 
You do not need to observe the Mpemba effect for
your experiment to be a success. Finding that the Mpemba
effect does not occur under certain conditions is still
a good experimental result. On the other hand, it is
certainly more dramatic and psychologically satisfying
if you can find conditions under which
the effect occurs. And if you find conditions under which it
occurs, you can then study what changes destroy the effect, which
provides a potentially valuable probe of the phenomenon.
You may want to do some preliminary
testing to find parameters where the Mpemba
effect occurs, and then decide how to proceed.

If you complete experiments on the Mpemba
effect, I'd be interested in hearing the results---if you
have a chance, send me an e-mail telling me what you found.
If enough experiments are done, perhaps this 2000+
year-old problem can be solved!

Finally, those who like the counterintuitive
nature of the Mpemba effect might be interested in
a similar phenomenon: water drops will last longer 
on a skillet well above $100^O C$, than on one only a little
above $100^O C$.
This is easier to explain
than the Mpemba effect~\cite{Liedenfrost}.

\acknowledgments{ We would like to thank Debra Waxman for 
providing a preliminary translation of the letter from Descartes 
to Mersenne, and Leigh Anne Eubanks for providing the final 
translation.  We would like to thank Nancy Ruff for
confirming the correctness of Burke's translation of Roger 
Bacon, over another, incorrect, published translation.  A 
previous version of this article appeared on the sci.physics 
FAQ web page~\cite{WebFAQ}.}



\begin{thebibliography}{99}
\bibitem{KuhnAssumptionNote} The explanation of ``heating
by antiperistasis'' is, of course, no longer valid.
But Kuhn is criticizing not just the explanation, but
the experimental results claimed by Marliani and Bacon.
\end{thebibliography}
\end{document}